\documentclass[10pt,a4paper,superscriptaddress]{revtex4}

\usepackage{graphicx}
\DeclareGraphicsExtensions{.pdf}

\usepackage{hyperref}

\begin{document}

\title{Time reversal symmetry in optics}

\author{Gerd Leuchs}
\email{gerd.leuchs@mpl.mpg.de}
\author{Markus Sondermann}
\email{markus.sondermann@physik.uni-erlangen.de}

\affiliation{Institute of Optics, Information and Photonics,
  University of Erlangen-Nuremberg, 91058 Erlangen, Germany}
\affiliation{Max Planck Institute for the Science of Light, 
  91058 Erlangen, Germany}

\begin{abstract}
The utilization of time reversal symmetry in designing and 
implementing (quantum) optical experiments has become more and 
more frequent over the past years.
We review the basic idea underlying time reversal methods, illustrate 
it with several examples and discuss a number of implications.\\
\textit{We dedicate this article to Professor Vladilen S. Letokhov, one of the
pioneers of laser spectroscopy, laser physics, quantum optics and
related fields. 
His many stimulating ideas influenced the development of these fields
and were often the starting point of novel sub fields flourishing ever
since. 
Very remarkably some of his ideas are still waiting to be
explored demonstrating that Professor Letokhov was well ahead of his
time.}
\end{abstract}

\maketitle

\section{Introduction}

Time reversal symmetry is a fundamental concept in physics.
Based on every day life observations, time reversal symmetry is far
from being obvious.
It often seems to be broken in that many evolutions apparently occur
only in one direction in time, \textit{i.e.}, they cannot run backwards.
In such cases irreversibility comes about, because the reverse
direction occurs only with a forbiddingly small probability.
One might call such processes thermodynamically irreversible but they
are certainly not irreversible in principle.
If we could control all degrees of freedom we would retrieve time
reversal symmetry.
Only in very special cases in particle physics there seems to be a
real violation of time reversal symmetry \cite{lueders1957}.
In all other cases time reversal symmetry is preserved and one may
take advantage of it.

Here, we concentrate on the field of optics and quantum optics.
Optics is governed by Maxwell's equations which obey time reversal
symmetry \cite{feynman1966}.
As outlined in one problem example below, already simple laboratory
tasks may be optimized based on this property.
But also in cutting edge problems of modern optics one can benefit
from taking guidance from time reversal symmetry.
One example is the efficient absorption of a single photon
by a single atom.
For perfect excitation of the atom the photon should have the shape of
the time reversed version of a spontaneously emitted photon.
This holds true in \emph{free space}
\cite{quabis2000,sondermann2007,stobinska2009}, 
in a waveguide \cite{rephaeli2010} as well as for atoms in a
resonator ~\cite{cirac1997}. 
We will discuss the free space absorption problem exemplary in more
depth below.
Time reversal symmetry arguments play a key role also in the storage
and retrieval of photons in atomic ensembles
\cite{moiseev2001,gorshkov2007}.

Methods based on time reversal symmetry have been applied successfully
to focusing of electromagnetic radiation onto non-quantum sources,
e.g., for microwave emitters embedded in a cavity
\cite{lerosey2004,carminati2007}.
Recently, sub-diffraction limited focusing has been demonstrated
in the optical domain in the absence of a source by focusing through a
scattering medium \cite{vellekoop2010}:
The wave front impinging onto a lens has been shaped such that it
resembles the phase conjugate, i.e., time reversed version of the
wave front that is generated by a hypothetical source emitting a wave
through the scattering medium. 
Another recent example of the application of time reversal techniques
is the determination of the optimal wave front for a new variant of
4$\pi$ microscopy \cite{mudry2010}.

The paper is outlined as follows:
In the next section, we highlight the relation between phase
conjugation and time reversal.
Then, the application of time reversal techniques in optics is
discussed for several different cases.
Finally, some issues related to phase conjugation of quantum states of
light are reviewed.

\section{Time reversal and phase conjugation}

\begin{figure}
\centerline{
\includegraphics{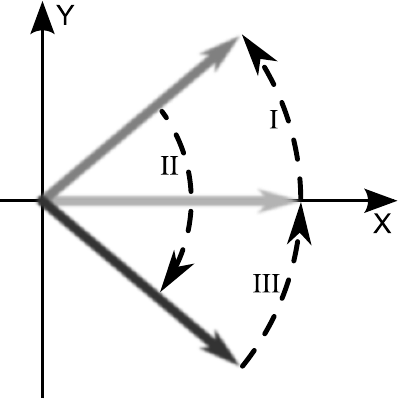}
}
\caption{\label{fig:phaseconjug} Most simple example for reversing an
  evolution in time: Phase space representation of the temporal
  evolution of a single mode optical field ($E(t)=X(t) + i Y(t)$) in
  its components. 
  (I): evolution starting at $t=0$ with oscillation $\exp(i\nu
  t)$. 
  (II): phase conjugation operation. 
  (III): continued evolution according to $\exp(i\nu t)$ back into the
  initial state. 
  The dynamics resembles the creation of spin echoes.} 
\end{figure}

We highlight the relation between time reversal and
phase conjugation (see, \textit{e.g.}, Ref. \cite{stenholm2004}).
For this purpose we express an electrical field $\vec{E}(\vec{r},t)$ by
  its spectral components $\vec{A}(\vec{r},\omega)$:
\begin{equation}
\label{eq:spectral}
\vec{E}(\vec{r},t)=\int_0^\infty\! d\omega \left[
\vec{A}(\vec{r},\omega)\mathrm{e}^{i\omega t} + 
\vec{A}^\star(\vec{r},\omega)\mathrm{e}^{-i\omega t}
\right]
\quad .
\end{equation}
Time reversal is equivalent to replacing $t$ by $-t$.
This results in 
\begin{equation}
\label{eq:spectral-conjug}
\vec{E}(\vec{r},-t)=\int_0^\infty\! d\omega \left[
\vec{A}(\vec{r},\omega)\mathrm{e}^{-i\omega t} + 
\vec{A}^\star(\vec{r},\omega)\mathrm{e}^{i\omega t}
\right] \quad.
\end{equation}
Thus, time reversal is equivalent to complex conjugating
all spectral amplitudes:
\begin{equation}
t \rightarrow -t \quad \Leftrightarrow \quad  
\vec{A}(\vec{r},\omega) \rightarrow \vec{A}^\star(\vec{r},\omega)
\quad .
\end{equation}
A graphical representation is given in Fig. \ref{fig:phaseconjug}.

\begin{figure}
\centerline{
  \includegraphics{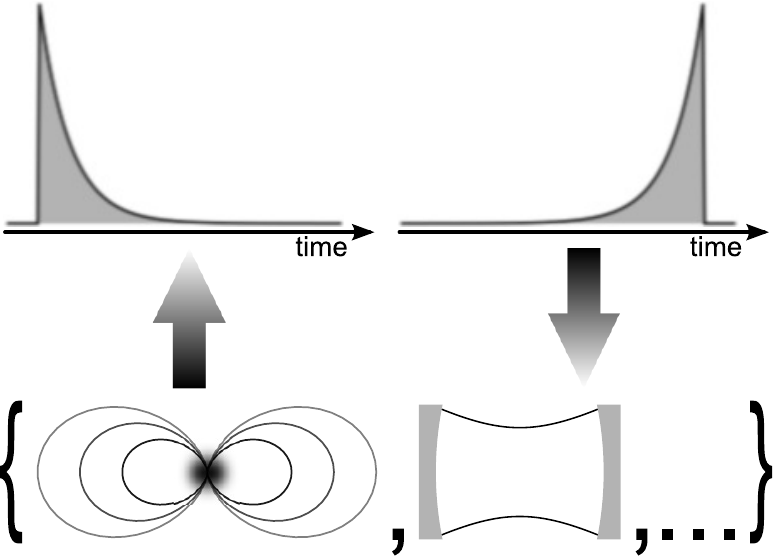}
}
\caption{\label{fig:exp_reversal} Paradigmatic systems involving
  exponentially decaying optical field amplitudes: spontaneous
  emission from a two-level atom with an electric dipole transition
  and field amplitude inside an empty optical resonator.} 
\end{figure}

We illustrate the equivalence stated above in a few examples.
Consider an electromagnetic wave with exponentially decreasing
amplitude, $E(t)\sim \exp(-\beta t + i\nu t)\cdot\theta(t)$,
$\theta(t)$ being the Heaviside function. 
The spectral amplitudes follow a Lorentzian distribution with 
$A(\omega)\sim [\beta+i(\omega-\nu)]^{-1}$.
Complex conjugation gives
$A^\star(\omega)\sim[\beta-i(\omega-\nu)]^{-1}$, which is the spectral
distribution of a wave with an exponentially increasing amplitude
$E(t)\sim\exp(+\beta t + i\nu t)\cdot\theta(-t)$.
The latter is clearly the time reversed version of the prior.
This example is relevant in the problem of exciting a single atom with
a single photon in different geometries
\cite{sondermann2007,stobinska2009,cirac1997,rephaeli2010} as well as
in the absorption of light pulses by an empty Fabry-Perot resonator
\cite{heugel2010} (see Fig. \ref{fig:exp_reversal} for illustration).

As an even simpler example consider a monochromatic (outward moving)
spherical wave $E(\vec{r})\sim\exp(-ikr+i\nu t)/r$.
Phase conjugating
$A(\vec{r},\omega)\sim\exp(-ikr)/r\cdot\delta(\omega-\nu)$ 
gives the amplitude distribution of a wave of same
frequency but moving inwards: $E(\vec{r})\sim\exp(ikr+i\nu t)/r$.
This illustrates that time reversal is equivalent to inversion of
motion in all degrees of freedom \cite{domingos1979}.

As a practical application, we mention the temporal inversion of
picosecond pulses with distorted temporal profile by phase conjugation
in a nonlinear medium \cite{joubert1989}.

\section{Examples in classical optics}

A simple example for the application of time reversal methods is
the coupling of light into a standard (single mode) optical fiber. 
One usually says that one has to mode match the beam which is to be
coupled to the optical mode of the fiber in order to achieve 100\%
coupling efficiency. 
However, this perfect mode matched light field is nothing else than
the time reversed version of the light field that is emitted from the
fiber end and collimated by the coupling lens.
Of course, upon time reversal the spatial profile remains the same and
for a monochromatic beam also the temporal profile is not altered.
The only change is in the direction of propagation.
Time reversal is equivalent to inversion of motion in all degrees of
freedom, i.e., the only difference with the wave originating from the
fiber is the opposite sign of the wave vector.
If the lens used for coupling exhibits aberrations the spatial profile
of the time reversed version must be shaped such that the aberrations
introduced by the lens are precompensated: the spatial phase
distribution is the complex conjugate of the wave collimated by the
lens.
This case is included in the spatial dependence of the amplitudes in
Eqs. \ref{eq:spectral} and \ref{eq:spectral-conjug}.

Another example is the coupling of light to a Fabry-Perot resonator.
What was stated for resonators with atoms inside in
Ref. \cite{cirac1997} is of course also true for an empty resonator.
So consider a resonator consisting of two perfectly reflecting
mirrors. Let a wave be traveling back and forth between the
two mirrors.
If at $t=0$ the reflectivity of one of the mirrors is suddenly
decreased light leaks out of the cavity. 
The field amplitude measured outside the cavity drops exponentially
with the cavity decay constant.
Upon time reversal, the light would propagate back into the resonator
with an exponentially increasing amplitude.
Therefore, the optimum light pulse that is coupled into an empty
cavity is an exponentially increasing one (see
Ref. \cite{heugel2010}).

An example for an apparent lack of time reversal symmetry is an
optical isolator (cf. \cite{bloembergen1989,tompkin1990}). 
Such a device is based upon the rotation of the polarization vector of
the electric field by the Faraday effect.
Typically an isolator consists of an entrance polarizer, the Faraday
rotator and an exit polarizer with its transmission axis rotated 45
degree away from the one of the entrance polarizer.
The strength of the magnetic field of the Faraday rotator is designed
such that it perfectly rotates the input polarization onto the axis of
the exit polarizer.
If now a beam is back reflected onto the exit polarizer with proper
state of polarization (\textit{i.e.}, the time reversed version of a wave
leaving the isolator), it passes the exit polarizer and is rotated by
45 degree by the Faraday rotator.
However, it is rotated such that it is blocked by the entrance
polarizer, because the polarization direction rotation due to the
Faraday effect is independent of the direction of light propagation. 
In the time reversal picture this is to be expected, since it is only
the evolution in the degrees of freedom of the wave that have been
reversed. 
Hence the current flow generating the magnetic field of the Faraday
rotator (which is of course also a degree of freedom) has also to be
reversed for recovering time reversal symmetry.

\section{A quantum optics example}

Next, we discuss a quantum optical example: the absorption of a single
photon  by a single atom in free space. 
The process of perfect absorption can be understood as the time
reversed version of spontaneous emission. 
Upon spontaneous emission from the excited atomic state, the
electromagnetic field (assumed to be in a vacuum state before the
emission of the photon) goes into the state of an outward moving
single photon dipole wave packet with an exponentially decaying
temporal mode profile. 
The time reversal argument now suggests that perfect absorption will
be achieved by an inward moving dipole wave \cite{quabis2000} with
exponentially increasing temporal profile \cite{sondermann2007}.
The latter requirement has been checked in a theoretical simulation
\cite{stobinska2009}, where the atom has been excited with a
one-photon Fock state which is a superposition of single
frequency mode Fock states with a Lorentzian weight.
The first requirement demands illumination from full solid angle,
which would require, e.g., an infinitely deep parabolic mirror as the 
focusing device.
Therefore, any finite focusing optics may enable absorption close to
but never exactly at 100\%.
The same holds true for the temporal domain, where in principle
infinitely long pulses are required.
However, a pulse length of about 4 atomic life times already boosts
absorption probabilities to 99.9\%  \cite{stobinska2009}. 

In practice, any focusing device may exhibit aberrations.
Since upon spontaneous emission the wave front emitted by the atom 
is disturbed by the aberrations of the redirecting optics, the time
reversed version must exhibit wave front aberrations of opposite sign. 
In other words, correcting for the aberrations of the focusing optics
is nothing else than making use of the equivalence between time
reversal and phase conjugation.

\begin{figure}
\centerline{
  \includegraphics{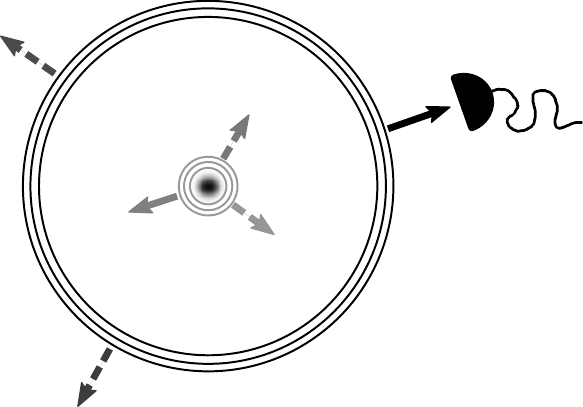}
}
\caption{\label{fig:momentum} The momentum of a spontaneously emitted
  photon and the center of mass motion of the atom left behind are
  entangled. Therefore, a detector measuring a photon with a
certain wave vector will project the atomic center of mass momentum
onto a certain direction.} 
\end{figure}

The arguments put forward so far are related to the optical degrees of
freedom.
For perfect absorption also the atomic degrees of freedom
have to be reversed.
In principle, upon emission of a photon, momentum conservation predicts
a recoil of the atom.
Before measuring or absorbing the photon, the  momenta of the photon
and the ground state atom are entangled (see Fig. \ref{fig:momentum}).
Using the time reversal argument one would have to generate an
incoming single photon dipole wave entangled with an incoming atomic
center of mass motion wave.
This clearly is a task beyond current technology.
Therefore, the idea is taking guidance from the M\"ossbauer effect:
tightly trapping the atom at one position in space in a volume well
within the Lamb-Dicke regime will effectively enlarge the mass of the
atom.
In the regime of quantized harmonic oscillation of the atom the photon
momentum will be transferred to the whole macroscopic trap, the
resulting motion of which is then negligible.
For suitable trap parameters see, \textit{e.g.},
Ref. \cite{leibfried2003}.
A detailed theoretical discussion of the recoil problem in
spontaneous emission can be found in Ref. \cite{chan2002}.

\section{Phase conjugation of quantum states of light}

In the example of the time reversed version of spontaneous emission
one might have the idea of creating the time reversed photon by direct
phase conjugation of the spontaneously emitted one.
While optical phase conjugation is technically demanding in the
classical domain one has to cope with its fundamental noise properties
in the quantum domain.

There are several implications when phase conjugating classical 
light fields. 
One is related to the spectrum of the light pulse to be phase
conjugated.
As has been shown by Ou, Bali and Mandel, the pulse shape reflected by
a phase conjugating mirror may become independent of the incident
pulse, if the incident pulse length is much shorter than the response
time of the phase conjugating mirror \cite{ou1989}.
This should not pose a problem.
The pulse length of a spontaneously emitted photon is
practically on the order of a few excited-state life times, which is
at least on the order of nanoseconds for most of the atomic dipole
transitions.
The response time of phase conjugating mirrors is considerably
shorter, as evidenced by the successful reversal of picosecond pulses
\cite{joubert1989}.

In contrast to that, the long pulse length does pose a technical
problem.
Pulse durations of several nanoseconds correspond to spatial pulse
lengths on the order of one meter.
For successful phase conjugation, the pulse should fit completely into
the conjugating material, which is possible in principle but not very 
practical given the above parameters.

A more serious obstacle has been put forward by Yamamoto and Haus
\cite{yamamoto1986} and by Gaeta and Boyd \cite{gaeta1988} who showed
that the phase conjugation process induces excess quantum noise.
This excess noise is related to the quantum noise limit of phase
insensitive optical amplifiers as explained in the next paragraph. 
Both, conjugation and amplification are non-unitary operations and can
only be implemented imperfectly, i.e. by introducing noise. 
Unitary operation can be retrieved by embedding the process in a
higher dimensional Hilbert space, i.e. by using additional, auxiliary
field modes \cite{leuchs2006AAMOP}.

One might be tempted to reduce the task of phase conjugating a light
field to implementing the transformation
$\hat{a}\rightarrow\hat{c}=a^\dagger$ much like an amplifier would
require $\hat{a}\rightarrow\hat{c}=\sqrt{G}a^\dagger$, $G$ being the
power gain.
However, neither of these operators fulfills the commutator relation
$[\hat{c},\hat{c}^\dagger]=1$.  
The problem is cured by allowing for
an auxiliary field mode $\hat{b}$ with $[\hat{a},\hat{b}]=0$.
The relations we are looking for are
described by the following Bogoliubov transformation: 
\begin{equation}
\hat{a}\rightarrow
\hat{c}=\gamma_1\hat{a}+\gamma_2\hat{a}^\dagger
+\gamma_3\hat{b}+\gamma_4\hat{b}^\dagger \quad .
\end{equation}
For phase conjugation $\gamma_2$ and  $\gamma_3$ have to be
different from zero, $\hat{a}\rightarrow\hat{c}=\gamma_2\hat{a}^\dagger
+\gamma_3\hat{b}$.
Using the field commutator for $\hat{c}$ one finds the relation 
$|\gamma_2|^2=|\gamma_3|^2-1$. 
Without loss of generality we assume that $\gamma_2$ and $\gamma_3$
are real, yielding $\gamma_3=\sqrt{\gamma_2^2+1}$. 

Some terms in the Hamilton operator for the total field,
$\hat{c}^\dagger\hat{c}+\frac{1}{2}$, do not preserve energy, one term
\textit{e.g.} leads to the creation of two photons, one each in mode
$\hat{a}$ and $\hat{b}$.
By adding a (quantized) pump field all terms in the Hamiltonian can
be modified to be energy conserving. 
But this step is not needed when the only purpose is estimating the
noise added by phase conjugation.  
Here we recall the relation between the variances for the amplitude
quadratures  
$X_i=\frac{1}{2}(\hat{i}+\hat{i}^\dagger)$, $i=a,b,c$ of the
input, auxiliary and output field, respectively, the variance being defined
as $\langle\Delta X_i^2\rangle=\langle X_i^2\rangle
-\langle X_i\rangle^2$.  
If the auxiliary field is in the ground state, the calculation
yields 
\begin{equation}
\langle\Delta X_c^2\rangle = \gamma_2^2\langle\Delta X_a^2\rangle
+ (\gamma_2^2+1)\langle\Delta X_b^2\rangle \quad .
\end{equation}
This quantifies the added noise. 
If the gain in the phase conjugating process is to be $\gamma_2^2=1$
phase conjugation adds two units of vacuum noise: 
\begin{equation}
\langle\Delta X_c^2\rangle = \langle\Delta X_a^2\rangle
+ 2\langle\Delta X_b^2\rangle \quad .
\end{equation}
This result was reported earlier by Caves \cite{caves1982} and Cerf
and Iblisdir \cite{cerf2001}. 
Thus one can say that there is no universal noiseless phase conjugation
much like there is no universal 'NOT' operation among the single qubit
gates \cite{gisin1999,buzek1999}.

However, given enough a priori information about a state, the phase
conjugated field may nevertheless be generated artificially with 100\%
fidelity.  
In conclusion, we underline the usefulness of the time reversal
symmetry concept as a powerful tool in designing optical experiments.

\end{document}